\documentstyle[epsfig,fleqn]{basi}
\begin{document}
\title[Dynamics of star clusters] 
      {On mass and velocity distributions in members of star clusters : An analytical approach}
\author[S. Chatterjee et al.]
       {S. Chatterjee,$^{1}$\thanks{e-mail: chat@iiap.ernet.in} Brijesh Kumar,$^{2}$ and Ram Sagar,$^{1, 2}$\\ 
        $^{1}$Indian Institute of Astrophysics, Bangalore 560 034, India \\
        $^{2}$State Observatory, Manora Peak, Nainital 263 129, India}
\maketitle
\label{firstpage}

\begin{abstract}
Dynamical evolution of stellar mass distribution in star clusters is analysed by considering simultaneously the effects of 
dynamical friction, stochastic heating and the gravitational potential due to mass distribution in the clusters. 
A simple expression is suggested for the dynamical friction which adequately describes it in both high and low velocity ranges 
and the effect of mass distribution in the cluster on the dynamics of a test mass can be described by an anharmonic 
potential. With the help of energy considerations we describe the dispersions in position and in velocity for stars of 
different masses showing how the slowing down and mass segregation of stars evolve in the cluster. Results are presented  
with the parameters for the star clusters of our Galaxy.
\end{abstract}


\section{Introduction}
Properties of star clusters form some of the most fascinating areas of astrophysics, particularly due to their close 
contact with issues related to star formation, mass function and stellar evolution. Such systems have stars over a wide range 
of masses ($m$) which lie in the range $0.1 \leq m/M_{\odot} < 10$. The 
distribution of these masses in real space as also in velocity space are signatures of various effects that arise in the context of 
dynamical evolution of the star in the gravitational potential of all the masses in the cluster. In what follows, we give a simple 
quantitative estimate of the evolution of velocity of any star of mass $M$, (for the moment only those whose life times are 
longer than the age of the cluster). The method is completely analytical and thus amenable to simple computational schemes. 
This is made possible firstly because a simple and yet accurate approximations for the Chandrasekhar dynamical
friction formula could be found and the same type of formulae also found to fit the gravitational field in the cluster. 
This choice enables us to integrate the equation of motion analytically and its usefulness 
with reference to stellar dynamics in a spherically symmetric potential is demonstrated in the subsequent sections with 
special reference to open clusters and globular clusters. The relaxation process is considered to be due to dynamical 
friction, the violent relaxation process being completed within few thousands of years. 
The mass burning upto this time since the birth of the star is indeed negligible 
and $v(0)$, $M(0)$ are thus the quantities that obtain after the completion of violent relaxation.
The gas and dust content in clusters being small, the drag forces due to dust and gas are neglected and only dynamical friction
due to stars is considered. As is known, the dynamical friction is
caused by the fluctuations in the trajectory of the star in question, due to encounters (not physical collisions) with other
stars present in the cluster. The other important parameter
that contributes to the dynamics of a test particle is the gravitational field arising out of the mass 
distribution (assumed Gaussian) in the cluster. We show that for a density profile $\rho(r)$ decaying with $r$, the 
gravitational potential $\phi(r)$ too can be adequately described by a function similar to that used to approximate the 
Chandrasekhar dynamical friction expression. A combination of these two approximations helps to integrate the equation 
of motion through simple mathematical operations. The integrated equation thus considers the dynamical evolution of 
a test mass $M$ due to combined effects of dynamical friction and its oscillation in an anharmonic $\phi(r)$ due to 
mass distribution in the cluster. It is indeed necessary to explain the motivation behind the 
present analytical work. This is mainly motivated by the recognition that "there is plenty of room for the development 
of new analytic techniques", "bringing into play more physically motivated reasoning in the choice of fitting 
functions" (Hut, 1996). The results that follow do give tractable analytical methods to estimate various physical processes 
in the evolution of a test mass in a cluster and may prove to be useful for the future theoretical understanding of the 
existing observational data. 

The main principles involved in the dynamics of the system 
are discussed in the next two sections. In section 2, we determine the effect of dynamical friction by considering stellar 
encounters. Essentially, the method is similar to what is given by Binney and Tremaine (1987) in which a King type 
velocity distribution has been incorporated. The main scheme for integrating the equation of 
motion is contained in sections 3 and 4 while the numerical scheme of solution is given in section 5. An interesting 
feature of our scheme is that it can easily accommodate modification due to mass loss arising out of various mechanisms that 
appear in the course of stellar evolution and is described in section 6. The effect of the above phenomena in 
describing mass segregation in the cluster is described in section 7 which is followed by results and discussions 
in section 8 of the paper.

\section{Dynamical friction with King type velocity distribution}
The role of gravitational interaction between the masses is considered in terms of stellar encounters. This, as is 
understood from the celebrated work of Chandrasekhar leads to (1) slowing down of the velocity of the particle in the 
direction of its initial velocity $\mbox{\boldmath$v$}_{\parallel}(0)$ and (2) enhancements of its 
velocity $\mbox{\boldmath$v$}_{\perp}$, in the 
component perpendicular to $\mbox{\boldmath$v$}_{\parallel}$. The role of interparticle interaction is incorporated through the 
encounters between the test mass $M$, travelling with velocity $\mbox{\boldmath$v$}_{M}$ and the scatterers $m$, travelling 
with velocity $\mbox{\boldmath$v$}_{m}$. As is well known, the inverse square nature of the interaction leads 
to a change $\Delta \mbox{\boldmath$v$}_{M}$, which is given by,

\begin{equation}
\Delta \mbox{\boldmath$v$}_{M} = \frac{2m|\mbox{\boldmath$V$}_{0}| \mbox{\boldmath$\alpha$}}{(M+m)} 
\left[ 1 + \frac{b^{2} V_{0}^{4}}{G^{2} (M+m)^{2}}  \right]^{-1/2}
\end{equation}

\noindent where $\mbox{\boldmath$V$}_{0} = \mbox{\boldmath$v$}_{m} - \mbox{\boldmath$v$}_{M}$ and the components of unit vector
$\mbox{\boldmath$\alpha$}$ in the direction of initial $\mbox{\boldmath$V$}_{0}$ is $\cos\,\theta_{0}$ and in direction 
perpendicular to $\mbox{\boldmath$V$}_{0}$ is $\sin\,\theta_{0}$, where $\cos\,\theta_{0}$ and $\sin\,\theta_{0}$ are given as 
$G(m+M)/\sqrt(G^{2} (m+M)^{2} + b^{2}V_{0}^{4})$ and $b^{2} V_{0}^{4}/\sqrt(G^{2} (m+M)^{2}+ b^{2}V_{0}^{4})$ 
respectively. These yield, 

\begin{equation}
|\Delta \mbox{\boldmath$v$}_{M \parallel}|^{2}  = \frac{4m^{2}V_{0}^{2}}{(m+M)^{2}} 
\left[ 1 + \frac{b^{2} V_{0}^{4}}{G^{2} (m+M)^{2}}  \right]^{-2}
\end{equation}
$$\hspace{-4.5cm} |\Delta \mbox{\boldmath$v$}_{M \perp}|^{2}  = \frac{4m^{2}V_{0}^{6}b^{2}}{G^{2}(m+M)^{4}} 
\left[ 1 + \frac{b^{2} V_{0}^{4}}{G^{2} (m+M)^{2}}  \right]^{-2} \eqno(2.1)$$

\noindent which gives,

$$\hspace{-5.0cm} |\Delta \mbox{\boldmath$v$}_{M}|^{2}  = \frac{4m^{2}V_{0}^{2}}{(m+M)^{2}} 
\left[ 1 + \frac{b^{2} V_{0}^{4}}{G^{2} (m+M)^{2}}  \right]^{-1} \eqno(2.2)$$

\noindent In order to determine the changes that occur in time $t$, we consider the encounter process to be 
uncorrelated and on an average $2 \pi n_{0} V_{0} b\,db$ encounters occur per unit time for particles passing within 
an interval $b$ and $b+db$. The physical situation puts the limits of $b$ as $b_{min} \leq b \leq b_{max}$ and 
the velocity distribution of $m$ follows a truncated Gaussian, 

\begin{equation}
 f(v_{m}) = \left\{ \begin{array}{ll}
                       \frac{An_{0}}{(2 \pi \sigma_{m}^{2})^{3/2}} 
                       \left[ e^{-v_{m}^{2}/2\sigma_{m}^{2}} - e^{-v_{e}^{2}/2\sigma_{m}^{2}} \right] & \mbox{for $v \leq v_{e}$}\\ 
                        0 & \mbox{for $v > v_{e}$}
                       \end{array}
              \right. 
\end{equation}

\noindent where $v_{e}$ is the escape velocity of the scatterers from the cluster and $A$ is a normalisation constant, 
being given by,

$$\hspace{-5.2cm}1/A = \left[ \mbox{erf}(x_{e}) - \frac{2x_{e}}{\sqrt{\pi}} e^{-x_{e}^{2}} - 
\frac{4}{3\sqrt{\pi}} x_{e}^{3} e^{-x_{e}^{2}}  \right] \eqno(3.1)$$

\noindent and we have defined $x=v_{m}/\sigma_{m} \sqrt{2}$ and $x_{e} = v_{e}/\sigma_{m} \sqrt{2}$.

\noindent The velocity distribution of the scatterers as given in equations (3, 3.1) indeed satisfies the consistency 
condition that the cluster contains only those stars with velocity which cannot escape from the potential 
well created by the mass distribution in the cluster. The estimate for $v_{e}$ can be easily made by integrating the 
Poisson equation and putting $v_{e}^{2}/2 = \phi(\infty)$, being thus consistent with the nature of the mass distribution. 
For an accurate determination of the dynamical friction one must keep in mind the mass spectrum and the mass 
dependence of the velocity dispersion $\sigma_{m}$. In the scheme presented below, we assume that the mass 
spectrum can be represented by an average mass $m$ and an average velocity dispersion $\sigma_{m}$.

\noindent It then follows from the equations (1), (2.2) and (3) that on account of encounters there appears a diminution 
of $\mbox{\boldmath$v$}_{\parallel}$ being given by, 

\begin{eqnarray*}
\frac{d\mbox{\boldmath$v$}_{M}}{dt} = 
\int_{0}^{\infty} \int_{b_{min}}^{b_{max}}\Delta \mbox{\boldmath$v$}_{m}\, 2 \pi b db\, V_{0}\, f(v_{m})\, 4\pi v_{m}^{2} dv_{m}
\end{eqnarray*} 
$$=-\frac{2 \pi G^{2} m (m+M)}{2\sqrt{2} \sigma_{m}^{3}} 
\left[ \ln(\chi_{max}/\chi_{min})\right] A \,n_{0}(m)\, \mbox{\boldmath$v$}_{M} F_{ch} (x)\eqno(4)$$ 

\begin{eqnarray*}
\frac{d|\Delta \mbox{\boldmath$v$}_{M}|^{2}}{dt} = \int_{0}^{\infty} \int_{b_{min}}^{b_{max}}
|\Delta \mbox{\boldmath$v$}_{M}|^{2}\, 2\pi bdb\, V_{0}\, f(v_{m})\, 4\pi v_{m}^{2} dv_{m}
\end{eqnarray*}
$$=\frac{4 \pi G^{2} m^{2}}{\sqrt{2} \sigma_{m}} 
\left[ \ln(\chi_{max}/\chi_{min})\right] A \,n_{0}(m)\, x^{2} F_{ch} (x)\eqno(4.1)$$ 

\noindent with $\chi \cong 1+ b^{2} \sigma_{m}^{2} / G^{2}(m+M)^{2}$ and for all practical purposes 
we shall approximate $\ln (\chi_{max}/\chi_{min}) \approx 2 \ln \Lambda $ with $ \Lambda = b_{max} / b_{min}$. The $F_{ch}(x)$
is given as 

$$F_{ch}(x) = \left\{ \begin{array}{ll} 
(\mbox{erf}(x) - \frac{2x}{\sqrt{\pi}} e^{-x^{2}} - 
\frac{4}{3\sqrt{\pi}} x^{3} e^{-x_{e}^{2}})/x^{3}& \mbox{for $x \leq x_{e}$}~~~~~~~~~~~~\\
(\mbox{erf}(x) - \frac{2x_{e}}{\sqrt{\pi}} e^{-x_{e}^{2}} - 
\frac{4}{3\sqrt{\pi}} x_{e}^{3} e^{-x_{e}^{2}})/x^{3}& \mbox{for $x > x_{e}$~~~~~~~~~~~~~~}
        \end{array} 
\right. \eqno(4.2)$$

\noindent It can be seen that the above equations converge to the Chandrasekhar formulae when $x_{e} \gg 1$. In the 
low velocity range the dynamical friction $ \propto \mbox{\boldmath$v$}_{M}$ 
and goes as $\mbox{\boldmath$v$}_{M}/|\mbox{\boldmath$v$}_{M}|^{3}$ for the high velocity 
range. Further, it can be seen from simple algebra that $v_{M}^{2}$ evolves as, 

\setcounter{equation}{4}
\begin{equation}
v_{M}^{2}(t+\delta t) = v_{M}^{2}(t) + 
2 \mbox{\boldmath$v$}_{M}(t) (d \mbox{\boldmath$v$}_{M}/dt)_{t} \delta t  + (d \langle v_{M}^{2}(t) \rangle/dt)_{t} \delta t  
\end{equation}

\noindent The second and third terms in equation (5) are consistent with the cooling term and heating term respectively in
Bekenstein and Maoz (1992) ( see equations 3.2.5 and 4.16, therein) and are also consistent with equations (161-163) in Nelson 
and Tremaine (1999). In a future work we plan to re-examine their works in the light of equations (9-11) in the present paper. 
From equations (4.1) and (4.2) it can be seen that the second term in (5) is always negative while the last term in (5) is 
positive. The former gives a slowing down due to dynamical friction being $ \propto m(m+M)$ while the latter gives 
stochastic heating and varies as $m^{2}$. Both originate in the inverse square interaction and act jointly, taking the 
system to equilibrium, being thus a manifestation of the fluctuation-dissipation theorem. For accurate evolution of the 
velocity $v_{M}^{2}$ we have to follow the formalism given in the next section. Full details of these calculations 
will be puplished shortly. We, however, present a summary in the next section in view of its importance on the question 
of evolution to equipartition. 

\subsection{On approach to equipartition}
The question concerns, how the kinetic energy $\frac{1}{2}Mv_{M}^{2}(t)$ evolves in presence of encounters. It can be shown 
that on a single encounter the change in the kinetic energy of the mass $M$ is given by (Binney and Tremaine 1987), 

\begin{equation}
\Delta (M v_{M}^{2}/2) 
= (M/2) \left[ 2 \mbox{\boldmath$v$}_{M}.\Delta \mbox{\boldmath$v$}_{M} + |\Delta \mbox{\boldmath$v$}_{M}|^{2} 
+ 2 \Delta \mbox{\boldmath$v$}_{M}. \mbox{\boldmath$v$}_{C} \right]
\end{equation}

\noindent where $\mbox{\boldmath$v$}_{M}$ is the change of velocity on encounter 
and $\mbox{\boldmath$v$}_{c} = (m\mbox{\boldmath$v$}_{m} + M\mbox{\boldmath$v$}_{M})/(m+M)$ is the velocity of the
center of mass of the system. Knowing the components $\Delta v_{\parallel}$ and $\Delta v_{\perp}$ as 
given in equations (1-3) we find the r.h.s. of equation (6) to be,

$$ \hspace{-2.5cm}-(M/2) \left[
 (v_{M \parallel}-v_{m \parallel}) 
 (Mv_{M \parallel}+mv_{m \parallel}) \frac{2G^{2}(m+M)}{G^{2}(m+M)^{2}+b^{2} V_{0}^{4}}\,\, + \right.$$
\begin{equation}
 \left.  (v_{M \parallel}-v_{m \parallel})^{3} (Mv_{M \perp}+mv_{m \perp}) 
\frac{2bG}{G^{2}(m+M)^{2}+b^{2} V_{0}^{4}} \right] 
\end{equation}

\noindent If we have a situation in which particles $m$ randomly encounter the test particle $M$ at values of $b$ 
which are $+$ve and $-$ve equally frequently, we have to use the 
averages $\langle b \rangle=0$ and  $\langle v_{M \parallel} v_{m \parallel} \rangle=0$ (e.g. $b$ $+$ve means that the projectile
approaches from above and it's track is bent downwards, while $b$ $-$ve means that it approaches from below and the track is
bent upwards, which essentially makes $\Delta v_{M\perp}$  to be $+$ve and $-$ve equally frequently, being thus zero on an 
average), so that 

\begin{equation}
\Delta(Mv_{M}^{2}/2) = 
-\left[Mm/(m+M)^{2}\right] \left[Mv_{M \parallel}^{2} - mv_{m \parallel}^{2}\right] 
\frac{2G^{2}(m+M)^{2}}{G^{2}(m+M)^{2} + b^{2} V_{0}^{2}}
\end{equation}

\noindent It shows that $Mv_{M}^{2}/2$ increases for $Mv_{M \parallel}^{2} < mv_{m \parallel}^{2}$ and decreases 
for $Mv_{M \parallel}^{2} > mv_{m \parallel}^{2}$, so that in the former case the kinetic energy of $M$ increases with 
scattering while it decreases in the latter case. Since there are $2 \pi b ~db~ V_{0} n f(v_{m})$ collisions per unit time, we find

\begin{equation}
(d/dt)(Mv_{M}^{2}/2)  = -2 \pi^{2} G^{2} m M (M v_{M \parallel}^{2} I_{1} - m I_{2}) n 
\end{equation}  

\noindent where

\begin{equation}
I_{1} = \int_{0}^{\infty} \frac{1}{V_{0}^{3}} 
\ln \left[ \frac{G^{2}(M+m)^{2} + b_{max}^{2} V_{0}^{4}}{G^{2}(M+m)^{2} + b_{min}^{2} V_{0}^{4}}\right] 
f(v_{m}) 4 \pi v_{m}^{2} dv_{m}
\end{equation}

\begin{equation}
I_{2} = \int_{0}^{\infty} \frac{v_{m \parallel}^{2}}{V_{0}^{3}} 
\ln \left[ \frac{G^{2}(M+m)^{2} + b_{max}^{2} V_{0}^{4}}{G^{2}(M+m)^{2} + b_{min}^{2} V_{0}^{4}}\right] 
f(v_{m}) 4 \pi v_{m}^{2} dv_{m}
\end{equation}

\noindent Dimensionally we note that $I_{2}/I_{1} \simeq \langle v_{m \parallel}^{2} \rangle$. Equation thus 
expresses the evolution of the kinetic 
energy $Mv_{M}^{2}/2$ of the test star and shows that its rate of change vanishes when the 
equipartition i.e. $M v_{M \parallel}^{2} = m \langle v_{m \parallel}^{2} \rangle$ is reached. 
Full details of this evolutionary process, 
particularly, with respect to $\mbox{\boldmath$v$}_{M}$ will be presented in a future publication.

It is to be noted that the Chandrasekhar formula of which equations (4.2) and (13.2) are variants 
actually considers the density to be 
uniform. We have applied here the same formalism to an $r$ dependent density profile in view of Kandrup's result 
(Kandrup 1981) that in an $r$ dependent density profile the Chandrasekhar formula can be adopted by using the local 
number density $n(\mbox{\boldmath$r$})$ for the average $n$, appearing in 
the Chandrasekhar formula . Thus on averaging over all possible 
orbits (that lie at different $r$' s) one may use the averaged number density, as we have done, to estimate the 
net effect of dynamical friction. Chandrasekhar's simplification in considering the initial $\mbox{\boldmath$v$}_{M}(0)$ 
and final $\mbox{\boldmath$v$}_{M}(\infty)$  to be straight lines have also been investigated and the corrections are found 
to be marginal (Kandrup 1983). Furthermore, Del Popolo and Gambera (1999) have shown that the numerical 
value of the dynamical friction can be changed from the Chandrasekhar formula if a power law 
dependence $n(r) \sim r^{-p}$ ($p > 0$ see equation (12) in this paper) be considered. We have , however, not incorporated 
this effect since such a power law dependence makes the number density $n(\mbox{\boldmath$r$})$ blow up 
for $r \rightarrow 0$. In the computed results, given below, all the parameters have been taken from well known observations.

\section{Principles and Approximations}
We consider a test mass $M$, with an initial velocity $v(0)$, undergoing motion in the cluster. This 
motion takes place in a gravitational 
field $\phi(\mbox{\boldmath $r$},t) = \phi_{0}(\mbox{\boldmath $r$}, t) + \delta\phi(\mbox {\boldmath $r$}, t)$ of which the 
first part describes slowly varying (in time) average gravitational 
field due to all stars in the cluster, while $\delta\phi(\mbox {\boldmath $r$}, t)$ is a rapidly varying time dependent potential 
that arises due to density fluctuations in the cluster owing to random velocities of 
stars i.e., $\delta\phi(\mbox {\boldmath $r$}, t)$ gives rise 
to dynamical friction. The equation of motion of the star consists of the equations $ d\mbox{\boldmath $L$}/dt = 
- \mbox {\boldmath$L$} \left[ \mbox{\boldmath$F$}_{dyn}(M,\mbox{\boldmath $v$})/   |\mbox{\boldmath $v$}|  \right] $ and  
$d L^{2}/dt= - 2 L^{2} \left[ F_{dyn} (M,\mbox{\boldmath $v$})/  |\mbox{\boldmath $v$}| \right] $ for the angular momenta and
 \"{r} $= -\partial \phi_{0} / \partial r + (L^{2}/m r^{3}) - \mbox{\boldmath $v$}_{r} 
F_{dyn}(M,\mbox{\boldmath $v$})/|\mbox{\boldmath$v$}|$ for the radial motion. In what follows we 
consider $L = 0$ i.e. a star with no angular momentum, oscillating radially across the cluster. The 
case for non zero $L$ will be dealt with separately in a future work. For the case at hand,  the equation of 
motion (we consider only the radial velocity) thus reads,

\begin{equation}
\frac{d \mbox{\boldmath $v$}}{dt}= - \mbox{\boldmath $\nabla$} \phi_{0}(\mbox{\boldmath $r$},t) -
\frac{\mbox{\boldmath $v$}}{\tau(M, \mbox{\boldmath $v$} )} = \mbox{\boldmath $F$}_{0}(r) 
          + \delta \mbox{\boldmath $F$}_{0}(\mbox{\boldmath $r$},t) - \mbox{\boldmath $F$}_{dyn}(M, \mbox{\boldmath $v$})
\end{equation} 

$$ \hspace{-7.7cm} \frac{d \mbox{\boldmath $v$}}{dt} \simeq - \mbox{\boldmath $\nabla$} \phi_{0}(\mbox{\boldmath $r$}) -
\frac{\mbox{\boldmath $v$}}{\tau(M, \mbox{\boldmath $v$} )} \eqno(12.1)$$ 

\noindent where the first term in the r.h.s. of equation (12) gives the steady gravitational potential field, while the second term
is a time dependent field arising due to changes in the density profile of the cluster in course of its dynamical evolution 
and the third term is the damping due to dynamical friction. By neglecting the 
$\delta \mbox{\boldmath $F$}_{0}(\mbox{\boldmath $r$},t)$ we obtain equation (12.1). 
The field $\delta \mbox{\boldmath $F$}_{0}(\mbox{\boldmath $r$},t)$ 
though time dependent is a slowly varying quantity as compared to 
$\nabla \delta \phi(\mbox{\boldmath $r$},t)$, which finally leads to dynamical friction. The net effect of the $\delta \phi (r,t)$
term has been employed here by introducing the last term i.e. dynamical friction in equations (12) and (12.1). We must note, 
however that the 
dynamical friction does have a role in the origin of the $\delta F_{0}(\mbox{\boldmath $r$},t)$ term also since the density 
profile evolves with time due to dynamical friction. Due to the slow variation 
of $\delta \mbox{\boldmath $F$}_{0}(\mbox{\boldmath $r$},t)$, this term is ignored in this zeroth order approximation, to be, 
however, considered in a more complete theory. 

For integrating  equation (12.1) two cases have to be considered separately. Equation 
(12.1) can be solved in its completeness by the method given by Van Kampen (1985). The method essentially involves
the identification of a smallness parameter $\epsilon$ and expansion of the solution $v$ as a power series in $\epsilon$. 
We find below that the system has a dimensionless parameter $\epsilon'$ = relaxation time/time period of oscillation. When 
relaxation are slow  we have $\epsilon = 1/\epsilon'$ and the star behaves as a damped oscillator. For fast relaxation 
$\epsilon'' \ll 1$ and one can identify $\epsilon = \epsilon'$ as the small parameter and the system behaves as an 
overdamped oscillator and hence does not demonstrate any to and fro motion. Considering the
mass distribution in the cluster to be $ \rho(r) = \rho_{0} \exp (-\gamma r^{2}) $ and a velocity distribution
as given in equation (5) we find,

$$ \hspace{-6.5cm} -F_{r}=\partial \phi_{0} / \partial r=\left(\pi^{3/2} G \rho_{0}/\gamma^{1/2}\right)y F(y) \eqno(13) $$  

\noindent where $y=r \sqrt{\gamma}$ and 

$$ \hspace{-3.3cm} \frac{1}{\tau(M,v,t)}
         = \left[\frac{4\pi G^{2} n_{0}m(m+M(0)-\Delta M(t)) \ln (\Lambda)}{2\sqrt{2}\sigma_{m}^{3}}\right] F_{ch}(x) \eqno(13.1)$$

\noindent where $F_{ch}(x)$ has been defined in (4.4), $\Delta M(t) =  |\int_{0}^{t} {(dM/dt) ~dt}| $ being 
the mass reduction upto the time $t$ after the birth of the star. It is clear that $\Delta M(t)$ depends upon the 
evolutionary phase of the star which is also determined by
its initial mass $M(0)$. In what follows we shall be concerned with the stars in the main sequence for which analytical 
expression for $\Delta M(0)$ will be used.
It is seen that for ${x \rightarrow 0}~~ F_{ch}(x) \sim 4/3\sqrt\pi$ while for ${x \rightarrow \infty}~~ F_{ch}(x) \sim 1/x^{3}$.
We find that $F_{ch}(x)$ can be approximated as 

$$\hspace{-9.5cm} F(x) = 1/[a+bx^{6}]^{1/2} \eqno(13.2)$$ 

\noindent with $a = 3.25$, $b = 1.28$ for globular cluster case and $a = 2.56$, $b = 4.09$ for open cluster case (see section 4) 
and , with a reliability of 99\%
as seen by Kolmogorov-Smirnov (KS) test, a simplification, which we shall use here for a very simple computational scheme.
This approximation for $F(x)$ also gives on integrating equation (13), and approximate expression for the gravitational potential 
inside the cluster, which reads 

$$\hspace{-10.0cm} \phi_{0}(r) = \phi_{0} \tilde{\phi}(y) \eqno(13.3)$$ where  

$$\hspace{-9.8cm} \phi_{0} = \pi^{3/2} G \rho_{0}/2\gamma \eqno(13.4)$$

$$\hspace{-9.0cm} \tilde{\phi}(y) = \int_{0}^{y} y' F(y') dy'  \eqno(13.5)$$ 

$$\hspace{-5.5cm} \tilde{\phi}(y) \simeq y^{2} F(y) + (3b/7) y^{7} [F(y)]^3 + .... + ....  \eqno(13.6)$$

We were not able to find a closed form equation for $\tilde{\phi}(y)$ and the expansion given in equation (13.6) is 
fairly accurate upto $y = 1.33$, which will correspond to $r \sim$ 40 pc for a typical globular 
cluster and $r \sim$ 13 pc for an open cluster with halo. We note on numerically integrating $\tilde{\phi}(y)$ 
that it is monotonically increasing function and $\tilde{\phi}(y) \rightarrow 1/2$ for $y \rightarrow \infty$. The 
escape velocity $v_{e}$ of particles is then given 
by $\frac{1}{2}v_{e}^{2} = \phi_{0} \tilde{\phi}(\infty)$ i.e. $v_{e}^{2} = \phi_{0}$ . In all the discussions that follow we are 
concerned with particles that are gravitationaly bound to the cluster and hence obey the restriction $v^{2} < v_{e}^{2}$. 
The choice of $\rho(r)$ as a Gaussian is indeed arbitrary but it does satisfy that $\rho(r)$ is peaked 
at $r = 0$ and falls off rapidly beyond a length $r_{0} \sim 1/\sqrt \gamma$ from the centre, which is an essential feature 
of the density profile of the cluster (Nilakshi et al. 2002). This results in a 
gravitational field $F_{r} \propto -r$ for $r \ll r_{0}$ 
and $F_{r} \propto -r^{-2}$ for $r \gg r_{0}$. In what follows we 
ascribe such a behaviour through equations (13) and (13.2) The choices $a$ and $b$ that we 
present here are the ideal ones for the 
Gaussian $\rho(r)$. However, for any other type of decay too similar approximations as given in equations (13) and (13.2) are 
possible in which 
the choices of $a$ and $b$ would be decided by the exact nature of the decay of $\rho(r)$ e.g. for an exponential density 
distribution $\rho(r) = \rho(0) e^{-\kappa r}$ we find that the $F_{0}(r)$ can be 
approximated to $F_{0} \simeq  (\kappa r) / [a_{1} + b_{1}(\kappa r) ^6] ^{1/2}$ with 90\% confidence as shown by 
the KS test. This observation thus demonstrates the usefulness of our approximation. For the subsequent calculation described 
here we continue to follow a Gaussian distribution for the density profile.

\section{Method of solution}
We note that the system introduces certain time scales. For $x \rightarrow 0, F(x) \rightarrow 1$, 
and for $x \rightarrow \infty, F(x) \sim (2/\sqrt\pi) (2/3) x^{3}$. Hence for low velocities we have a relaxational 
time scale $\tau(M)$, 
$ 1/\tau(M,\mbox{\boldmath $v$})= (4\pi G^{2} n_{0}m(m+M) \ln (\Lambda)/2\sqrt{2}\sigma_{m}^{3}) (4/3\sqrt \pi)$ and an  
oscillational time period $T_{0}$ which is akin to the crossing time such that $ 4\pi^{2}/T_{0}^{2} = (4\pi/3) G \rho_{0}$. 
The solution of equation (12.1) must consider two distinct cases. ({\it a}) $T_{0} \ll \tau(M)$ and ({\it b}) $T_{0} \gg \tau(M)$. 

\subsection{Case {\it (a)} : $T_{0} \ll \tau(M)$; $\epsilon' = \tau(M)/T_{0}$; $\epsilon = 1/\epsilon'$ }
We examine the problem in two cases. Firstly we consider a typical globular cluster 
with $n_{0}$ = 25 pc$^{-3}$, $m=1M_{\odot}$, $\sigma_{m}$ = 6 km s$^{-1}$, $r_{0} = 1/\sqrt \gamma \sim$ 30 pc 
with $v_{e} = 22$ km s$^{-1}$. Next we consider
the intermediate age open cluster NGC 2099 (M37) for 
which $n_{0} \approx$ 3 pc$^{-3}$, $m=1M_{\odot}$ and $\sigma_{m} \approx$ 1 km s$^{-1}$, $r_{0} = 1/\sqrt \gamma \sim$ 10 pc, 
being the range upto which the halo is present. For this last case we find $v_{e} = 2.6$ km s$^{-1}$ 
(Kalirai et al. 2001, Nilakshi and Sagar 2002). 
For stars with $M(0) \leq 10M_{\odot}$ one finds $\tau_{relax} = 2\tau(M(0),0) > 9.2 \times 10^{7} $ 
years while  $T_{osc} \approx 2.75 \times 10^{7} $ years. Since $ \tau_{relax}/T_{osc} \approx 3$, the star does 
exhibit oscillations,
$r(t) = a(t) \cos(2\pi t /T_{osc})$ where $r(t)$ is the instantaneous radial distance of the star from the center of the cluster.
Indeed the nature of $F_{r}$ shows the potential $\phi_{0}(r)$ to be anharmonic, this anharmonicity can be incorporated in 
the calculation  by making the oscillational time scales $T(n)$'s to be dependent on the amplitudes of the oscillation for the 
respective cycles $n$ of oscillations of the particle. Indeed the complete dynamics requires the inclusion of higher harmonics 
of $2\pi/T(n)$ but as far as the energy loss is concerned it is the slowest oscillational mode that outweighs the 
contribution from other faster modes, due to rapid elimination of fast variables (Van Kampen 1985). It thus 
suffices to approximate the motion only in terms of the slowest $2\pi/T(n)$ mode as is done below.
It is to be noted that with the passage of time the amplitude of oscillation $a(t)$ will decay due to dynamical friction. 
$T_{osc}$ is also a time dependent quantity arising due to anharmonicity of the 
potential $\phi_{0}(r)$. If however, $\tau_{relax}/T_{osc} < 1$ the system becomes overdamped and such oscillations 
are not to be found as will be seen in case (b).  

The energy loss per cycle due to dynamical friction is then found by noting that power loss is the product of the velocity
and the resistive force, so that on defining $E_{n}$ to be the energy of the particle in its n$^{th}$ cycle, 

\begin{eqnarray*}
E_{n+1} - E_{n} = \Delta E_{n} = - \int_{0}^{T(n)} {F_{dyn}(M(t), v) ~v ~dt} 
\end{eqnarray*}
$$ = -2 \sigma_{m}^{2} X_{o}^{2} \beta \int_{0}^{T(n)} 
{\frac{(m + M_{\odot}-\Delta M~t)~x_{n}^{2} \cos^{2} (2\pi t /T(n))} {[1+x_{n}^{6} \cos^{6} (2\pi t / T(n))]^{1/2}} dt} \eqno(14)$$

\noindent where $X_{0}^{2} = (a/b)^{1/3}$, $\beta = 4\pi G^{2} n_{0} m \ln (\Lambda) /2\sqrt{2} \sigma_{m}^{3}/\sqrt a$,
and $x_{n}^{2} = v_{n}^{2}/ 2\sigma_{m}^{2} X_{0}^{2}$, $v_{n}$ being the amplitude of the velocity in the nth cycle and 
$T(n)$ is the corresponding time period. The above difference equation can be written as

\setcounter{equation}{14} 
\begin{equation}
z_{n+1}-z_{n}= -\frac{\beta}{2} \frac{z_{n}} {[1+z_{n}^{6}]^{1/2}} (m+M_{\odot}-\Delta M(t)) T(n) \frac{F_{N}(z_{n})}{F_{D}(z_{n})}
\end{equation}

$$\hspace{-4.0cm} F_{N}(z_{n})= 
1-\frac{5}{32} \frac{z_{n}^{6}}{[1+z_{n}^{6}]^{1/2}} + \frac{1965}{2097152} \frac{z_{n}^{12}}{[1+z_{n}^{6}]^{1/2}} \eqno {(15.1)}$$ 

$$\hspace{-7.2cm} F_{D}(z_{n})= 1 - \frac{1}{10} \left(\frac{T_{0} \sigma_{m}}{R}\right)^{2} z_{n}^{2} \eqno {(15.2)} $$ 

\noindent where $z_{n}^{2} = (48/15)^{1/3}x_{n}^{2}$, $t= \sum T(n)$, $R= (\gamma/2)^{-1/2}$ = average radius of the cluster,
$(2\pi/T_{0})^{2}=(4\pi/3)G\rho_{0}$, $T_{0}$ being the 
oscillation time period of the star in the harmonic limit and $t$ is the total
time elapsed after the formation of the star. In equation (15) $F_{N}(z_{n})$ denotes diminution of dynamical friction
as the star acquires larger velocity as it passes through the center of the cluster, while $F_{D}(z_{n})$ as given in
equation (15.2) gives the increase 
of the period of oscillation on account of anharmonicity. By defining,

\begin{eqnarray}
F[z(t),z(0)] = \sqrt{(1+z^{6}(t))} - \sqrt{(1+z^{6}(0))} - \sinh ^{-1} (1/z^{3}(t)) +  \sinh ^{-1} (1/z^{3}(0)) 
\end{eqnarray}

\noindent Equation (15) can be finally integrated to yield (by writing $t=nT_{0}$), 

$$ \hspace{-1.1cm} F[z(t),z(0)][F_{D}(z)/F_{N}(z)] 
          = -(3/2) \beta \left(m~t + M(0) ~t - \int_{0}^{t}{\Delta M(t')dt'}\right) \eqno(16.1)$$

\noindent where $z(t)$ thus denotes  the variation of $z$ when change over is made from discrete $n$ to continous 
variable $t$. Using the 
above equation (16) we plot $v(t)$ with $t$ for globular clusters and open clusters. 

As a first step we solve equation (16) by taking as a zeroth order approximation $F_{N} = 1 = F_{D}$. In the next step 
we solve equation (16) again by taking $F_{N}(z)=F_{N}(z0(t))$ and $F_{D}(z)=F_{D}(z0(t))$. 
The resulting $z(t)$ values are next used to 
calculate $F_{N}(z)$ and $F_{D}(z)$, and solve for the new $z(t)$ by using these 
values of  $F_{N}(z)$ and $F_{D}(z)$ in equation (16).  
This iterative scheme of calculation is now being extended to take care of iteration upto any desired level. Also the scheme can
take care of the effects of mass loss if the rate of mass loss ($dM/dt$) can be specified for the star in different stages of
its evolution. 

\subsection{Case {\it (b)} : $T_{0} \gg \tau(M)$; $\epsilon' = T_{0}/\tau(M)$; $\epsilon=\epsilon'$}
This is a case when the star relaxes within a single oscillational time period. This is indeed an overdamped case. Here
we take only the second term in equation (12.1) as the zeroth order term while the first term is a perturbation.
This is easily integrated in the zeroth order approximation to be,

\begin{eqnarray*}
F[z(t),z(0)]=\sqrt{(1+z^{6}(t))} - \sqrt{(1+z^{6}(0))} + \sinh ^{-1} (1/z^{3}(t)) -  \sinh ^{-1} (1/z^{3}(0)) 
\end{eqnarray*}
$$ = -3 \beta \left(m~t + M(0)~t - \int_{0}^{t}{\Delta M(t')dt'}\right) \eqno(17) $$  

\noindent Since in case (b) we have discarded the role of harmonic potential, it describes the motion for a free particle under 
damping due to dynamical friction. Comparison of equation (16.1) with equation (17) shows 
that the factor 3/2 on the r.h.s. of equation (16.1) is 
replaced by 3 in equation (17). This 
is because when we deal with a particle in absence of the harmonic potential its variation in 
velocity is due to damping alone. For a particle in a harmonic (also anharmonic potential) the oscillation 
makes $v$ to oscillate and the dynamical friction also tunes itself accordingly. In effect the damping is faster in case (b)  

The above equation (17) can be easily solved numerically.
This is a solution which gives $v(t) \sim v(0) \exp(-t/\tau(m))$ for $v \ll \sigma_{m}$ and $[v(t)^{3} - v(0)^{3}] = -3\beta t$
for $v \gg \sigma_{m}$ as is known for motion under dynamical friction, in absence of a potential.  Next we define

\setcounter{equation}{17}
\begin{equation}
r(t) = \int_{0}^{t}{v(t') dt'}
\end{equation}

\noindent which can be calculated from solution of $v(t)$ as given in equation (17). We now use the numerical values of $r(t)$ 
for the first term in equation (12.1), being thus treated as a perturbation. We note that the role for the first term 
in equation (12.1) is to provide an extra acceleration:

\begin{equation}
F_{0}(t) = (\pi^{3/2}G\rho_{0}/\gamma^{1/2})~~(\gamma^{1/2}r(t))~~ F(\gamma^{1/2}r(t))  
\end{equation}

\noindent We now add this extra term to equation (12.1) and integrate. This leads to 

\begin{eqnarray*}
F[z(t),z(0)]=\sqrt{(1+z^{6}(t))} - \sqrt{(1+z^{6}(0))} - \sinh ^{-1} (1/z^{3}(t)) +  \sinh ^{-1} (1/z^{3}(0)) 
\end{eqnarray*}
$$=-3 \beta \left(m~t+M(0)~t - \int_{0}^{t}
{\Delta M(t')dt'}\right)-(1/\sqrt 2 \sigma_{m} X_{0}) \int_{0}^{t} {F_{0}(t) dt} \eqno(20)$$ 

\noindent where the last term in equations (20) describes the effects of the gravitational field and is calculated from the 
solution of equations (17) and using the solution to numerically evaluate equations (18, 19) and the second term 
in equation (20). The r.h.s. of equation (20) being thus known in its entirety, the same numerical method used for 
solution of equation (17) can be repeated to find the solution of equation (20). 

\section{Numerical scheme of solution}
Time development of the velocity of the star in the potential of all other stars is contained in the 
dimensionless quantity $z(t)$ which has to be solved from equation (16) when $ T_{0} \ll \tau(M)$ and 
from equations (17) and (20) for
$T_{0} \gg \tau(M)$. It is , however, necessary to note that in both cases the formal schemes are the same. Essentially, 
we have to solve equations of the type,

\setcounter{equation}{20}
\begin{equation}
f(z(t),z(0)) = \chi(t)
\end{equation} 

\noindent In case (a) $ f[z(t),z(0)] = F[z(t),z(0)] [F_{D}(z) / F_{N}(z)]$ while in case (b) f[z(t),z(0)] = F[z(t),z(0)] and 
the $\chi(t)$ are the r.h.s. of equations (16), (17) or (20) as the case may be. The numerical scheme of solution is
thus essentially the same and is described below.

Given any initial condition $z(0)$ we tabulate the l.h.s. of equation (21) for different $z$ and $\chi(t)$ is also caculated for 
different $t$. It is then easy to find graphically the point $z(t)$ at which $f[z(t),z(0)]$ equals $\chi(t)$ which is shown in
figure 1. One must, however, observe that while evaluation of the l.h.s. of equation (21) is a simple numerical work, evaluation
of the r.h.s. of equation (21) needs inputs on mass loss from various sources. 

\begin{figure}[t]
\centerline{\psfig{file=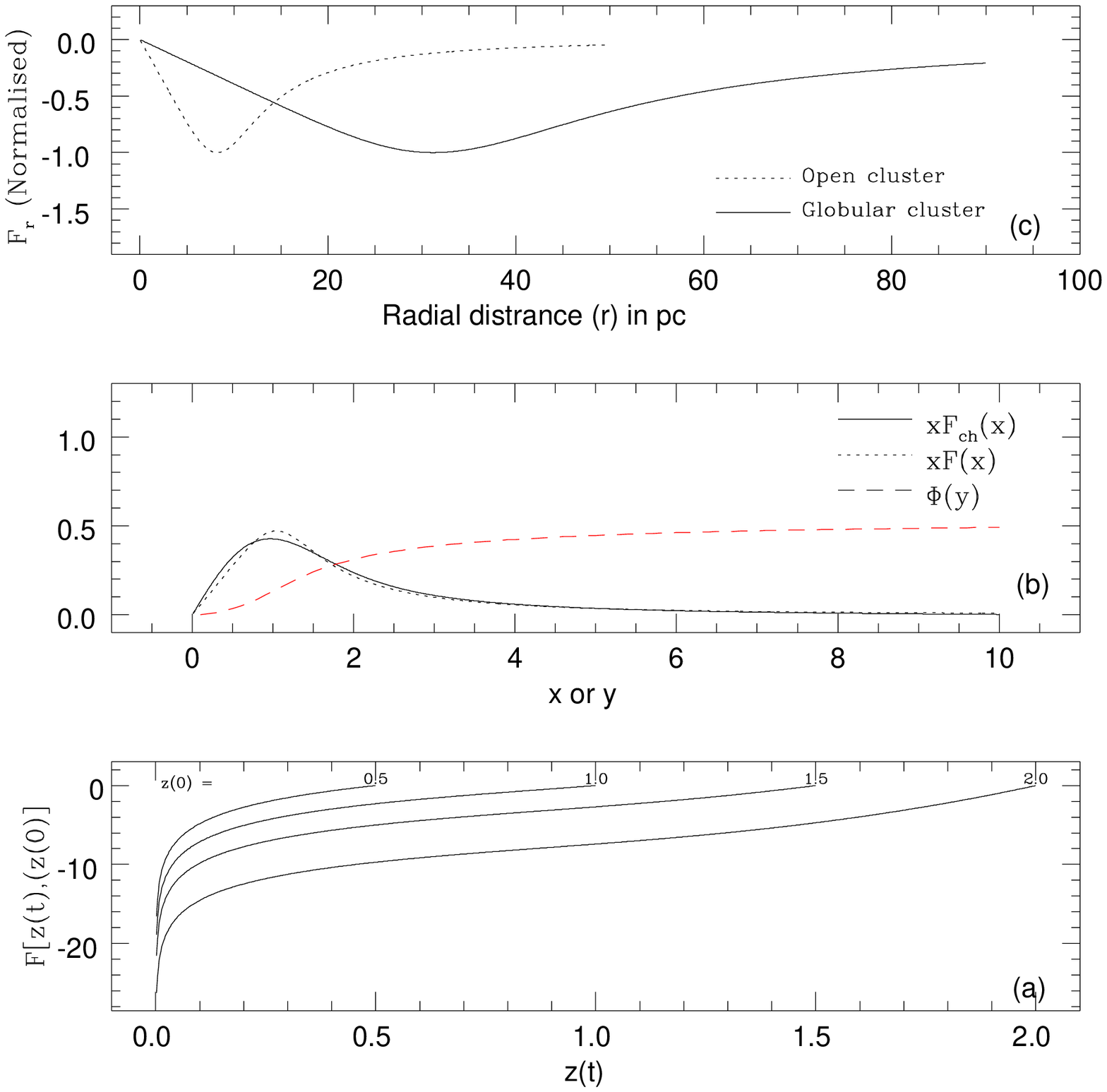,height=10cm,width=10cm}}
{{\bf Figure 1.} (a) F[z(t),z(0)] for different z(0) values, (b) 
Approximations for $xF_{ch}(x)$,$xF(x)$, $\Phi(y)$ for the globular cluster case (c) Normalised 
gravitational field in the open cluster (NGC 2099) and a typical globular cluster.}
\end{figure}

\section{Calculation of $\Delta M(t)$}
We assume $t$ to begin from ths star formation stage and $\Delta M(t)$ must contain the mass loss from all the 
sources e.g. mass burning, winds, mass ejection etc. A notable feature of the present theory is that the mass appears only 
on the r.h.s. of the equations (16), (17) and (20), while the dynamical variables like $v$ appear only on the l.h.s. This 
greatly helps the numerical work and yet the evolutionary history and dynamical history of the particle are both 
accommodated in one single equation. It is clear that the mass reduction in the main sequence occurs due to mass burning
of upto 10\% of core hydrogen, converting only a fraction of 0.0007 of the total hydrogen mass into energy. The quantity
$\int_{0}^{t} \Delta M(t') dt'$ 
due to mass burning can thus be neglected in comparison to $M(0)t$, within the main sequence life
time of the star. In this phase however, some mass is also lost due to stellar winds, the total mass loss due to
mass burning and stellar winds is taken from Schaller (1992) for solar composition and the 
quantity $\int_{0}^{t} \Delta M(t') dt'$ can thus be computed. This goes as an input to the r.h.s. of equations (16, 16.1, 17, 20).
Beyond the main sequence life time, the mass loss $\Delta M(t)$ becomes extremely complex for analytical 
calculation (Wilson 2000, Kudritzki and Puls 2000, Iben and Renzini 1983, Iben 1974).  However the $\Delta M(t)$ values 
provided by Schaller (1992) incorporates all these processes for stars in various phases of their 
evolution and has been incorporated to find $\int_{0}^{t} \Delta M(t') dt'$ in our computation and thus to find the dynamical 
evolution of stars of different masses present in the cluster. 

\begin{figure}[t]
\centerline{\psfig{file=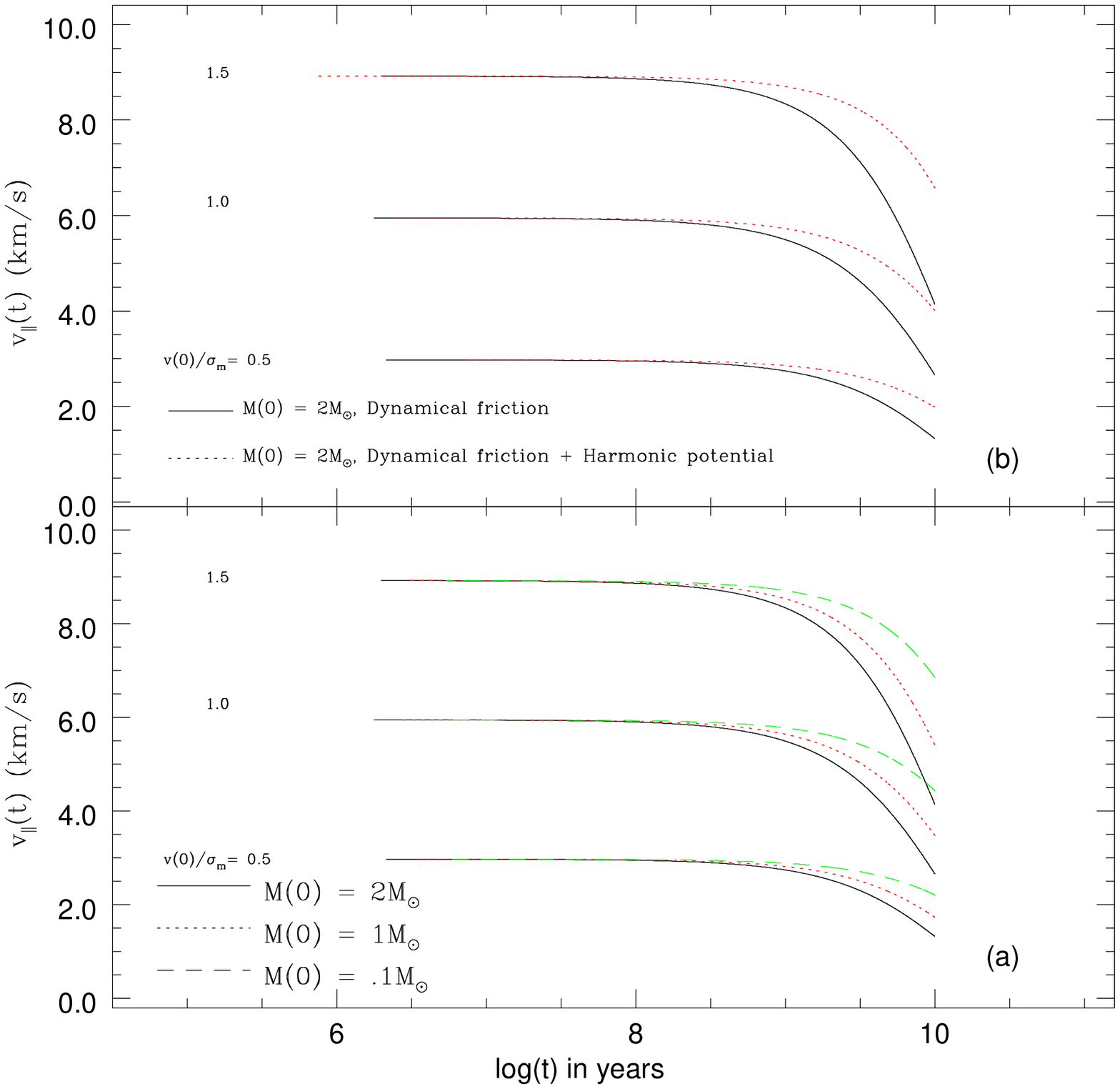,height=10cm}}
{{\bf Figure 2.} (a) Evolution of $v_{\parallel}$ for different masses at different initial velocities under dynamical friction 
                alone. (b) Effect of dynamical friction on the evolution of velocities under harmonic potential.}
\end{figure}

\section{On the question of mass segregation} 
The role of dynamical friction is to slow down fast moving particles and to finally equilibrate the system so 
that the velocities of the different masses reach their respective equipartition values. This energy loss by the test 
mass will result in a gradual shrinking of the amplitude $a(t)$ of its oscillation inside the cluster. This thus results in 
a segregation of the heavier mass towards the center of the cluster. How the mass segregation evolves with time can be 
measured by noting the variation of $\langle r(t) \rangle$ with $t$ from the virial theorem. By applying the virial theorem, 
$2\langle T \rangle = \langle -r \partial V / \partial r \rangle$, we find, (on noting 
that $2\langle T \rangle = 2 . \frac{1}{2} \langle v^{2} \rangle = 2. \frac{1}{2} \langle \frac{1}{2} v_{n}^{2} \rangle $, where 
averaging $\langle \cdots \rangle$ is done over a cycle),  

\begin{equation}
v^{2}/2 = (\pi^{3/2} G \rho_{0} / \gamma) 
\left[ \frac{\gamma \langle r^{2}\rangle}{[a + b \gamma^{3} \langle r^{2} \rangle^{3}]^{1/2}}\right]
\end{equation}

\noindent where $v^{2}$ (valid for $v^{2} < v_{e}^{2}$ ) is calculated from equations (14-16)
and $\langle r \rangle \sim \langle r^{2} \rangle ^{1/2}$. The variation of $\langle r \rangle$ with $t$ for 
different masses $M$ is a question of importance in studying the mass segregation in the cluster.

\begin{figure}[t]
\centerline{\psfig{file=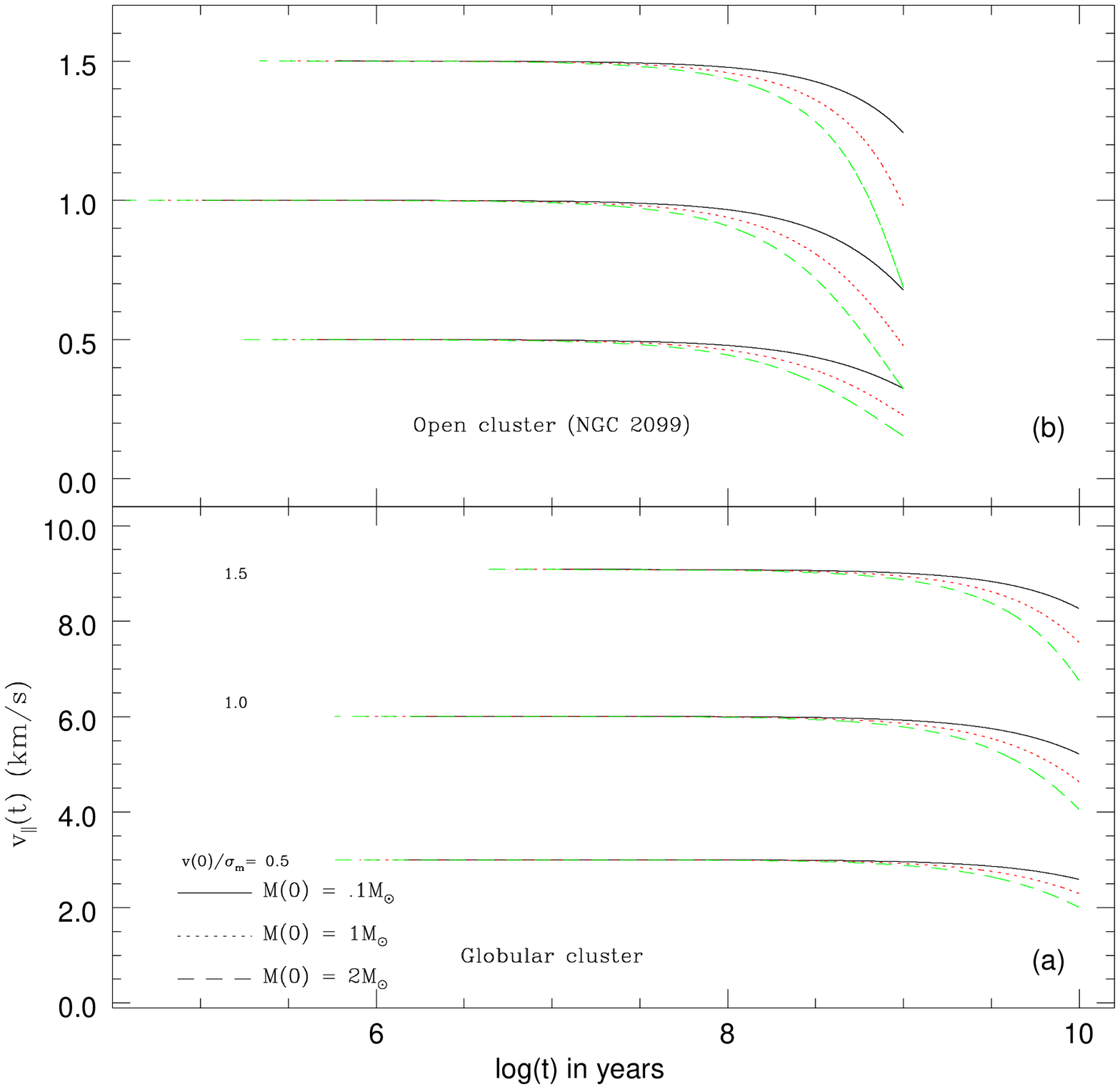,height=10cm}}
{{\bf Figure 3.} Evolution of velocities for a globular cluster (a), an open cluster (b) The captions of panel (a) applies 
                to (b) also. }
\end{figure}

\section{Results and discussions}
Investigations contained in the present paper give a method to study the dynamics of a star in a potential due
to a spherically symmetric distribution of masses, - a situation that arises in star clusters - $v_{\parallel}$ of the star 
being damped by the dynamical friction with $v_{\perp}$ being enhanced by the stochastic heating. We have considered 
the mass distribution in the cluster to be falling off in a Guassian fashion with the distance $r$ from the center of 
the star cluster. Such a variation is chosen because close to the center of the cluster, the gravitational potential 
is expected to be harmonic, while the anharmonic part takes over only at longer distances. To calculate the 
dynamical friction, the Chandrasekhar type formula which uses King type distribution of velocities of stars has been used. 
However this modified Chandrasekhar formula being quite 
involved, we have approximated it by a simple power law type of expression (13.2), which has an excellent fit with the 
modified Chandrasekhar formula, as seen from Kolmogorov-Smirnov test. It is indeed this identification of a simple formula 
which fits
the Chandrasekhar formula very satisfactorily - that makes our numerical scheme very simple, fast and easy to check and
yet accurate. The 
computational scheme is further facilitated by the fact that it is adequate to approximate the gravitational field $F_{r}$ by
equation (13) and the gravitational potential $\phi(r)$ by equations (13.3 - 13.6). The results that follow are based on 
the above inputs and implies that we are considering an epoch after the completion of the violent relaxation phase, 
at which an initial mass segregation described by $\rho(r)$ has taken place. We have chosen $b_{min} =$ average interparticle 
distance and $b_{max} =\sqrt(3/2\gamma)$. 

The results of our calculations are found in figures 1-6, which are indeed self explanatory. In our computations we 
have used the numbers given in section 4 of the paper. The method of solution is illustrated in in figure 1(a) 
where we show the plot of $F[z(t),z(0)]$ with $z(t)$ for various initial values $z(0)$. As a check we have tried 
to see the evolution of velocity of a test particle in absence of the gravitational potential which is essentially the solution 
of equation (17).  We find that in such a case the velocity decay 
follows $v_{\parallel}(t) = v_{\parallel}(0) exp(-t/\tau(m))$ for $v_{\parallel} \ll \sigma_{m}$ 
and $[v_{\parallel}(t)]^{3} = [v_{\parallel}(0)]^{3} - 3\beta t $ for $v_{\parallel} \gg \sigma_{m}$ (for 
low $t$ values)  as is expected from the Chandrasekhar dynamical friction. This is illustrated in figure 2(a), where only
the effect of dynamical friction is retained. The same scheme of solution when applied to equation (17) yield the result in 
figure 2(b), where we find that due to oscillations in the potential well, the dynamical friction gets modulated in terms of 
instantaneous velocity and hence the decay may be slower. The usual mass dependences are also borne out due to the dependence of
dynamical friction on the mass $M$ of the projectile. This also has bearing with the fact that lower masses lose small amounts of
their mechanical energy, while larger masses lose large amounts of their initial mechanical energy. This has been 
adequately described by equations (9-11). However, we cannot give at present accurate $v$ dependences of the 
integrals in (10) and (11) and hence the exact nature of approach towards equipartition cannot be described in the 
present paper and is reserved for a future work. An approximate scheme is contained in (5), which takes
care of dynamical friction as well as stochastic acceleration\footnote{Why equation (5) cannot describe approach 
to equipartition is because the two averages used in calculating $\langle dv_{\parallel}/dt\rangle$ 
and $d \langle (\Delta v)^{2}\rangle/dt$ have been considered to be independent. An averaging subject to the constraint (8) 
will lead to equipartition and will be discussed in a future paper.}. It shows the following qualitative 
behaviour that the decrement in $v^{2}(t)$ is more pronounced for the higher mass stars. This, indeed results from a 
competition between the two terms in equation (5) which shows that the net loss of energy due to dynamical friction 
goes as $m(m+M)$ while the stochastic acceleration - in effect a diffusion in velocity space - goes 
as $m^{2}$ (Chandrasekhar 1943, Spitzer and Schwarzschild 1951, Wielen 1977). For a 
more exact result in a realistic situation these effects with proper mass spectrum are to be considered (Chatterjee 1991, 1995). 
This is illustrated in figure 4. These processes
will also describe $\delta F_{0}(r,t)$, which we have ignored in our present approximation. It can indeed be incorporated as a 
perturbation once we understand the effects of condensation and evaporation from our present work.

The change of $v^{2}(t)$ for different masses, enables us to make contacts with earlier 
workers (e.g. Gorti and Bhatt 1996, 1996a, who consider the dynamical effects of protostellar clumps in gas clouds and 
Aarseth and Heggie 1998). Their numerical works essentially contain the description of the evolution process as seen 
in equation (5). This evolution of $v^{2}(t)$ also serves as the means to investigate the mass segregation in clusters. 
Given the value of $v^{2}$, the value of $\langle r \rangle \sim \langle r^{2} \rangle ^{1/2}$ for a star of given 
mass $M$ can be found from equation (22), the results 
being represented in figures 5 and 6. In obtaining these results we have considered only heavier 
stars i.e. $1 \leq M/M_{\odot} \leq 5$ i.e. $m \ll M$. In obtaining 
figure 6 we have noted that the heavier stars evolve to their late phase in $\sim 10^{8}$ years 
i.e. within a time scale in which very little effect is observed with respect to their dynamical evolution.
The subsequent scenarios of their evolution being not known with precise models, we have considered these stars to evolve only
dynamically and noted their condensation with time, through the parameter $\langle r(t)\rangle$. In other words figure 6 suggests 
the "tentative" locations within which moderately heavy stars or their remnants e.g. neutron stars are to be found as they evolve
between $10^{5}$ - $10^{10}$ years.  

\begin{figure}[t]
\centerline{\psfig{file=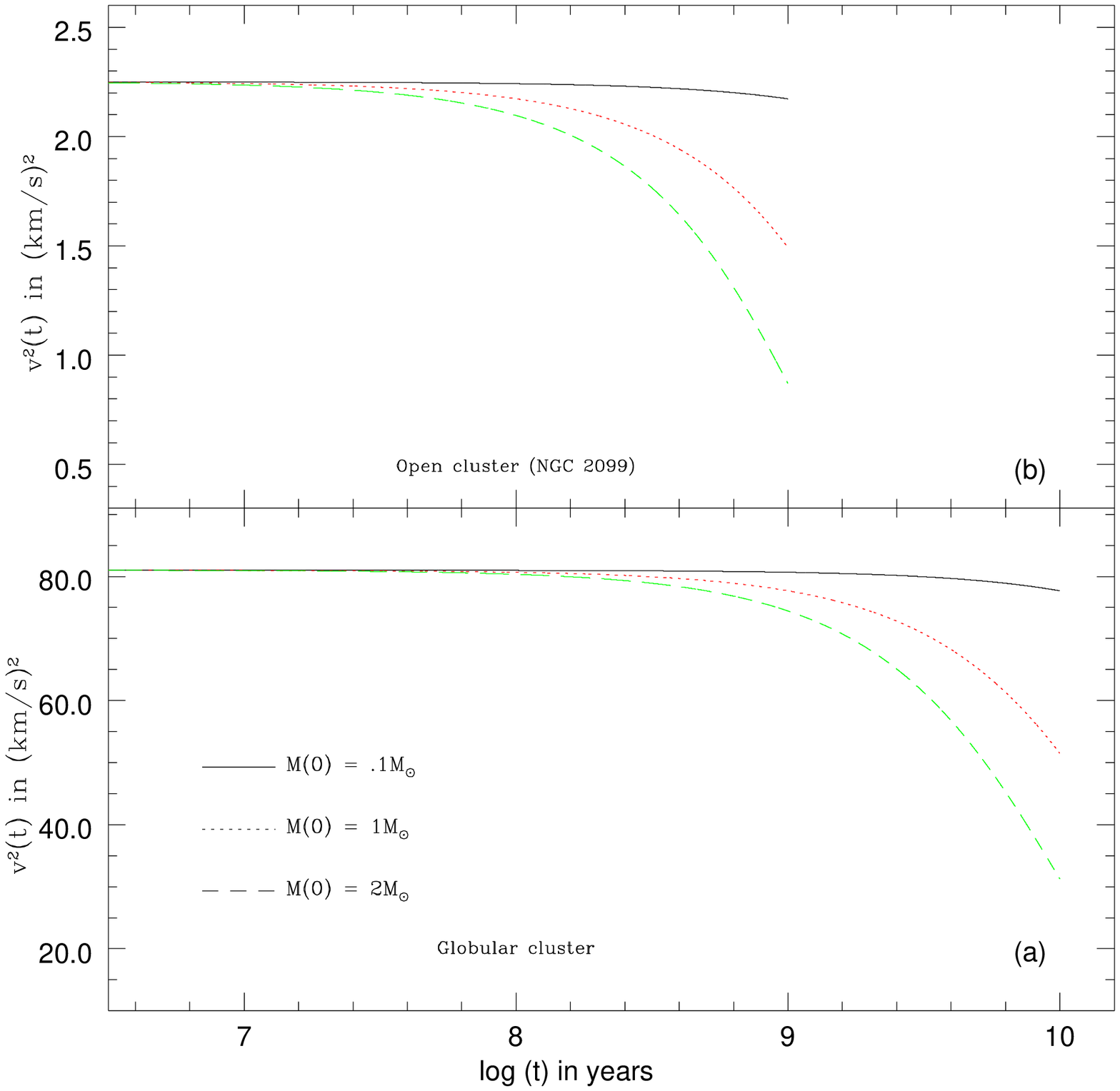,height=10cm}}
{{\bf Figure 4.} Change in $v^{2}(t)$ for different masses when $v(0) = 1.5 \sigma_{m}$.}
\end{figure}

We return again to figure 5 and 6, to understand the possible origin of an extended corona in open 
clusters (Nilakshi et al. 2002, Pandey et al. 2001, Durgapal and Pandey 2001). It is seen that 
if we have a $1M_{\odot}$ star with $v(0) = 1.5\sigma_{m}$, it can occupy upto 11 pc initially (newly born) in an open 
cluster, while occupy only upto 5 pc after $10^{9}$ years. The corresponding numbers for a typical globular cluster 
are 21 pc at birth and 20 pc at the end of
$10^{10}$ years. If on the other hand we consider a $0.1M_{\odot}$ star with $v(0) = 1.5 \sigma_{m}$ at birth, it will
continue to live within 11 pc if it be in a typical open cluster and within 21 pc if it be in a typical globular cluster, 
because of very weak velocity loss due to feebleness of the dynamical friction that it experiences. It is thus possible for such 
a star to travel far away from the center of the open cluster in view the smallness of the gravitational field that arises in an
open cluster. This may thus explain the presence of an extended corona in older open clusters. However, if the corona are
integral part of star formation process itself as has been indicated by Nilakshi et al. (2002), then the dynamical evolution
will only modify it.   

\begin{figure}[t]
\centerline{\psfig{file=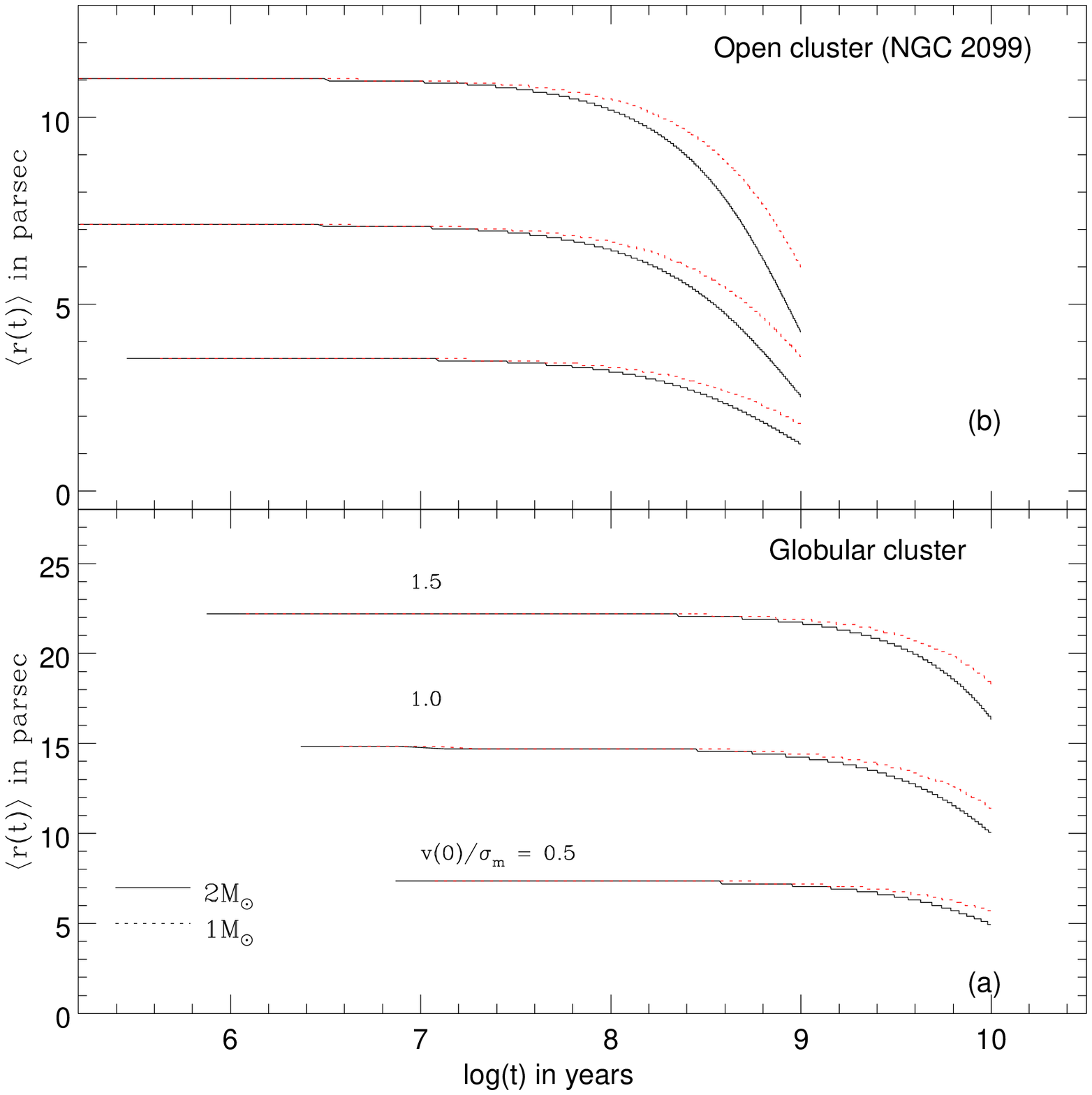,height=10cm}}
{{\bf Figure 5.} $\langle r(t)\rangle $ versus $t$ for a globular cluster (a) and  an open cluster (b). The captions of 
                                panel (a) applies to (b) also.}
\end{figure}

The competition of dynamical friction with stochastic heating as given in equation (5) is indeed qualitatively correct when 
dealing with $M \gg m$ but always leads to a decrease in $v^{2}(M)$ for all masses. The correct approach should 
be to obtain $v^{2}(M)$ from equation (11), which slows down the evolution of $v^{2}(M)$ as it 
approaches the equipartition value. This approach as also equation (5) in the present paper do predict the situation
that leads to the lighter mass  (a) lying very close to the edge of the corona or (b) 
escaping the gravitational potential of the cluster. Concerning case (a) we note that this has been borne out in several recent 
works (Durgapal and Pandey 2001, Pandey et al. 2001, Nilakshi et al. 2002) who find a long 
tail for the corona, owing to the presence of low mass stars
. Though the density profile $\rho(r)$ determined by them can not be fitted to a Gaussian curve and definitive form is 
not yet established, the long tail in the corona, can still be considered to give rise to gravitational field of the type
 $x F(x)$ given by equation (13.3), which we have already tested by considering various types of $\rho(r)$'s with the coefficients
$a$ and $b$ of the fit being decided by the nature of the $\rho(r)$. The dynamics that can follow from such a fit is in the
process of being tested by us and in a preliminary study is seen to be consistent with the models
concluded through observations.
To estimate evaporation process we ought to solve the Boltzmann equation by using the appropriate scattering cross section
for the $\sigma (v_{\parallel}, v_{\perp})$ an inverse square field (Chandrasekhar 1943, 
Spitzer and Schwarzschild 1951), but for a particle bound in an anharmonic
potential $\phi(r)$ given by equation (13.1). Such a method will at one stroke combine (1) dynamical friction (2) stochastic 
heating and (3) diffusion in $r$ space in an anharmonic potential $\phi(r)$. The last mentioned has been qualitatively addressed 
here but a full scale understanding (e.g. of the density profile and role of evaporation in describing the 
density profile) does deserve such a 
combined attack. As far as we are aware, this problem has not been tackled till now, to which we will
return shortly along with the role of an initial mass function and velocity dispersion on the dynamical evolution of a cluster. 

\begin{figure}[t]
\centerline{\psfig{file=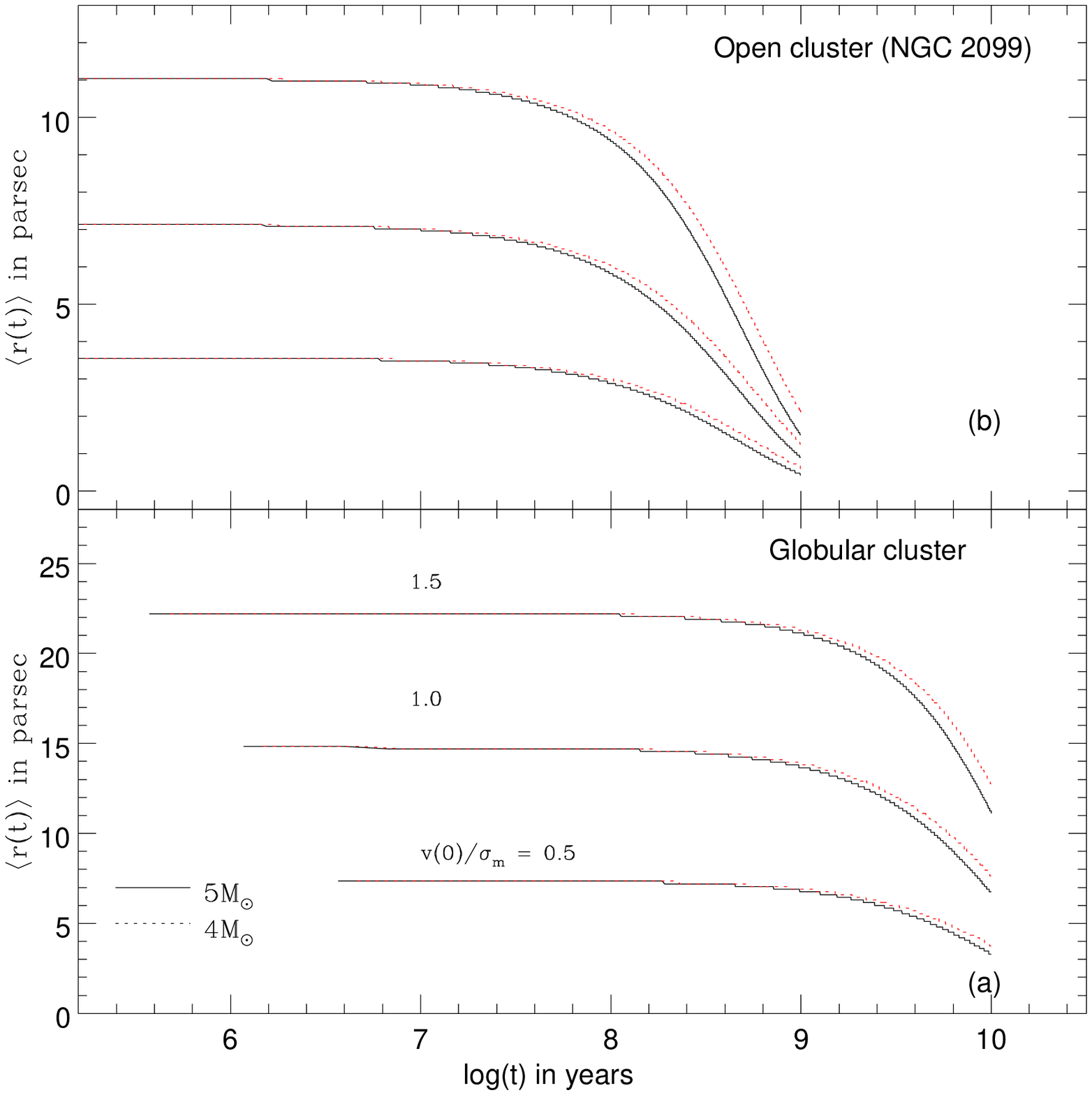,height=10cm}}
{{\bf Figure 6.} $\langle r(t)\rangle $ versus $t$ for a globular cluster (a) and  an open cluster (b). The captions of 
                         panel (a) applies to (b) also.}
\end{figure}

\section*{Acknowledgements}
The authors would like to thank Professor H.C. Bhatt for useful comments on an earlier version of the 
manuscript. They thank Dr. A.K. Pandey for his comments on several features of open clusters, as seen by recent observations.
One of the authors (SC) would like to thank Professor D.C.V. Mallik for his encouragement.
The referee of the paper is thanked for useful comments.

\label{lastpage}
\end{document}